\documentclass{aa}
\usepackage{epsf,epsfig}
\begin{document}

\title{More evidence for hidden spiral and bar features in bright 
early-type dwarf galaxies
\thanks{Based on observations collected at the European Southern Observatory,
Chile}}

\author{Fabio D. Barazza \inst{1}
\and Bruno Binggeli \inst{1} 
\and Helmut Jerjen \inst{2}} 

\offprints{F.D. Barazza, e-mail : barazza@astro.unibas.ch}   

\institute{Astronomisches Institut, Universit\"at Basel, Venusstrasse 7,
CH-4102 Binningen, Switzerland
\and
Research School of Astronomy and Astrophysics, The Australian National
University, Mt Stromlo Observatory, Cotter Road, Weston ACT 2611, Australia}

\date{Received 25 February 2002 / Accepted 31 May 2002}

\abstract{Following the discovery of spiral structure in IC3328 (Jerjen et
al.~2000), we present further evidence that a sizable fraction
of bright early-type
dwarfs in the Virgo cluster
are genuine disk galaxies, or are hosting a disk component.
Among a sample of 23 nucleated dwarf
ellipticals and dS0s observed with the Very Large Telescope in $B$ and $R$, we
found another four systems exhibiting non-axisymmetric
structures, such as a bar and/or spiral arms, indicative of a
disk (IC0783, IC3349, NGC4431, IC3468). Particularly remarkable are the
two-armed spiral pattern in IC0783 and the bar and trailing arms in NGC4431.
For both galaxies the disk nature has recently been confirmed by a rotation
velocity measurement (Simien \& Prugniel 2002). Our photometric
search is based on a Fourier decomposition method and
a specific version of unsharp
masking. Some ``early-type'' dwarfs in the Virgo cluster seem to be
former late-type galaxies which were transformed to 
early-type morphology, e.g.~by ``harassment'', 
during their infall to the cluster, while
maintaining part of their disk structure.
}
\maketitle

\section{Introduction}
The physical nature and origin of early-type dwarf (essentially dE) galaxies
is still largely unknown (for a review see Ferguson \& Binggeli 1994).
According to their apparent flattening, which is similar to the one measured
for giant ellipticals (Binggeli \& Popescu 1995; Ryden \& Terndrup
1994), early-type dwarfs seem to be spheroids. This is supported by the fact
that most of these systems are not rotation-supported (Ferguson \& Binggeli
1994; de Rijcke et al. 2001; Geha et al. 2001). On the other hand, it has 
always been suspected that a considerable number of early-type
dwarfs might be disk galaxies. In fact, around 20 bright early-type 
dwarfs in the Virgo cluster are classified dS0 because of their S0-like 
morphology, exhibiting a lens, a bar, or high flattening (Sandage \&
Binggeli 1984; Binggeli \& Cameron 1991). In certain evolutionary scenarios,
like ram-pressure stripping (Gunn \& Gott 1972; Abadi et al. 1999) 
or galaxy harassment (Moore et al. 1998), dwarf ellipticals are believed 
to have
originated from late-type spirals or irregulars, hence some of them might have
retained their disk nature. In addition, Ryden et al. (1999) showed that
many early-type dwarfs have ``disky'' isophotes, similar to giant ellipticals.

Recently, Jerjen et al.~(2000) discovered a weak spiral structure in the
seemingly normal dwarf elliptical galaxy IC3328 by means of deep
VLT-photometry. This is unambiguous evidence for the presence of a disk in 
this galaxy, supporting the conjecture that the number of (hidden) disk
galaxies among bright early-type dwarfs could be quite high.

Following the work of Jerjen et al. (2000), we carefully searched
a larger sample of dEs observed with the VLT for additional indications of
spiral or bar structure. We first applied the
same techniques as for IC3328, i.e.~relying on residual images and
Fourier analysis. Among the 23 dEs studied
we found seven promising candidates for hidden disk structure. 
However, when we further explored 
these objects we realized that the observed spiral and bar features which
we took as disk signatures
are accompanied, and could actually be {\em caused}, by a specific
behaviour of the
ellipticity and position angle profiles.
By performing a set of simulations with artificial galaxies we convinced
ourselves that the interplay
between photometric parameters can indeed produce 
amazingly spiral-like twisting isophotes and thus mimic a genuine spiral 
structure. Hence, although 
the particular parameter combinations found in many galaxies
are remarkable in themselves, 
they cannot unambiguously be
interpreted as signs of disk structure. 

Fortunately, the ambiguity can be solved 
by applying an unsharp masking technique, which is a model-free
method to amplify {\em local}\/ image residuals. Only four of the seven
candidates mentioned (two of which are
already classified as dS0), in addition to IC3328 (Jerjen et al.~2000),
withstood the scrutiny of unsharp masking: IC0783, IC3349, NGC4431, and
IC3468. These are the galaxies focused on in the present paper.
Although none of the four discovered cases is as spectacular as 
IC3328 with its tightly wound spiral, they show that the fraction of disk
galaxies among bright early-type dwarfs, at least in the Virgo cluster,
is 20\% or larger. We recall that the situation is quite similar to 
classical (non-dwarf) elliptical galaxies, where also some Es
turned out to be barred S0s, or to contain such a component, as betrayed
by inner isophotal twists (Nieto et al.~1992).
\begin{table*}[t]
\caption[]{Basic data of the early-type dwarfs considered in this study}
\vspace{0.3 cm}
\begin{center}
\begin{tabular}{lllrrrrrr}
\hline
\multicolumn{1}{l}{VCC} &
\multicolumn{1}{l}{Name} &
\multicolumn{1}{l}{Type} &
\multicolumn{1}{c}{$R_T$} &
\multicolumn{1}{c}{$M_{R_T}$} &
\multicolumn{1}{l}{$B-R$} &
\multicolumn{1}{c}{$v_{\odot}$} &
\multicolumn{1}{l}{$\epsilon$} &
\multicolumn{1}{c}{$pa$} \\
\multicolumn{1}{l}{(1)} &
\multicolumn{1}{l}{(2)} &
\multicolumn{1}{l}{(3)} &
\multicolumn{1}{c}{(4)} &
\multicolumn{1}{c}{(5)} &
\multicolumn{1}{c}{(6)} &
\multicolumn{1}{c}{(7)} &
\multicolumn{1}{c}{(8)} &
\multicolumn{1}{c}{(9)} \\
\hline
\\
0490 & IC0783  & dS0(3),N  & 12.63 & -18.52 & 1.34 & 1293 & 0.25 & 130 \\
0940 & IC3349  & dE1,N     & 13.56 & -17.59 & 1.25 & 1563 & 0.22 &  14 \\
1010 & NGC4431 & dS0(5),N  & 12.47 & -18.68 & 1.39 &  913 & 0.38 & 168 \\
1422 & IC3468  & E1,N:     & 12.64 & -18.51 & 1.16 & 1372 & 0.17 & 150 \\
\\
\hline
\end{tabular}
\end{center}
\end{table*}

The rest of the paper is organized as follows. In Sect.~2 we give some
observational background and put this investigation into context 
with our more general project. 
In Sect.~3 we employ, and present the results of, a Fourier
analysis of the galaxy images. Sect.~4 contains a 
brief account of the application of unsharp masking,
which is the decisive detection tool. In Sect.~5 we discuss our
findings and what they could mean, case by case. A summary is given 
in Sect.~6. Throughout this
paper we assume a distance to the Virgo cluster of $D = 17$ Mpc, corresponding
to $(m-M) = 31.15$.

\section{Observational background}
This work is part of a larger project aimed at the determination of distances
to dwarf elliptical 
galaxies in the Virgo and Fornax clusters by means of the Surface
Brightness Fluctuations method, and a detailed photometric analysis of their
brightness distributions. The galaxies were primarily selected by their
morphological appearance, i.e.~early-type dwarfs (dE, dS0), and by their
apparent size, i.e.~an isophotal radius $r_{B,25}\,>\,30''$. So far, 25 objects
have been observed in two runs at the Very Large Telescope at ESO Paranal
Observatory in service mode. Details of the observations are to be reported
elsewhere (Jerjen et al., in preparation). The important parameters
of the four galaxies considered in this study are listed in Table 1. 
The columns of the Table are as follows:
{\em columns} (1) and (2): identification of the observed galaxy;
{\em column} (3): morphological type in the classification system of
Sandage \& Binggeli (1984), taken from Binggeli et al.~(1985);
{\em column} (4): total $R$-band magnitude, corrected for galactic 
extinction (from Barazza et al., in preparation);
{\em column} (5): absolute $R$-band magnitude;
{\em column} (6): total colour index $B-R$; 
{\em column} (7): heliocentric radial velocity in km\,s$^{-1}$
(from the NED);
{\em columns} (8) and (9): ellipticity $\epsilon =
1-\frac{b}{a}$, where $a$ and $b$ are the major and minor axis,
and position angle $pa$ (from top
counterclockwise), respectively, determined at an isophotal level of
$\sim 25 {\rm mag}/\sq\arcsec$ in $R$.  

\begin{figure*}[t]
\begin{center}
\epsfig{file=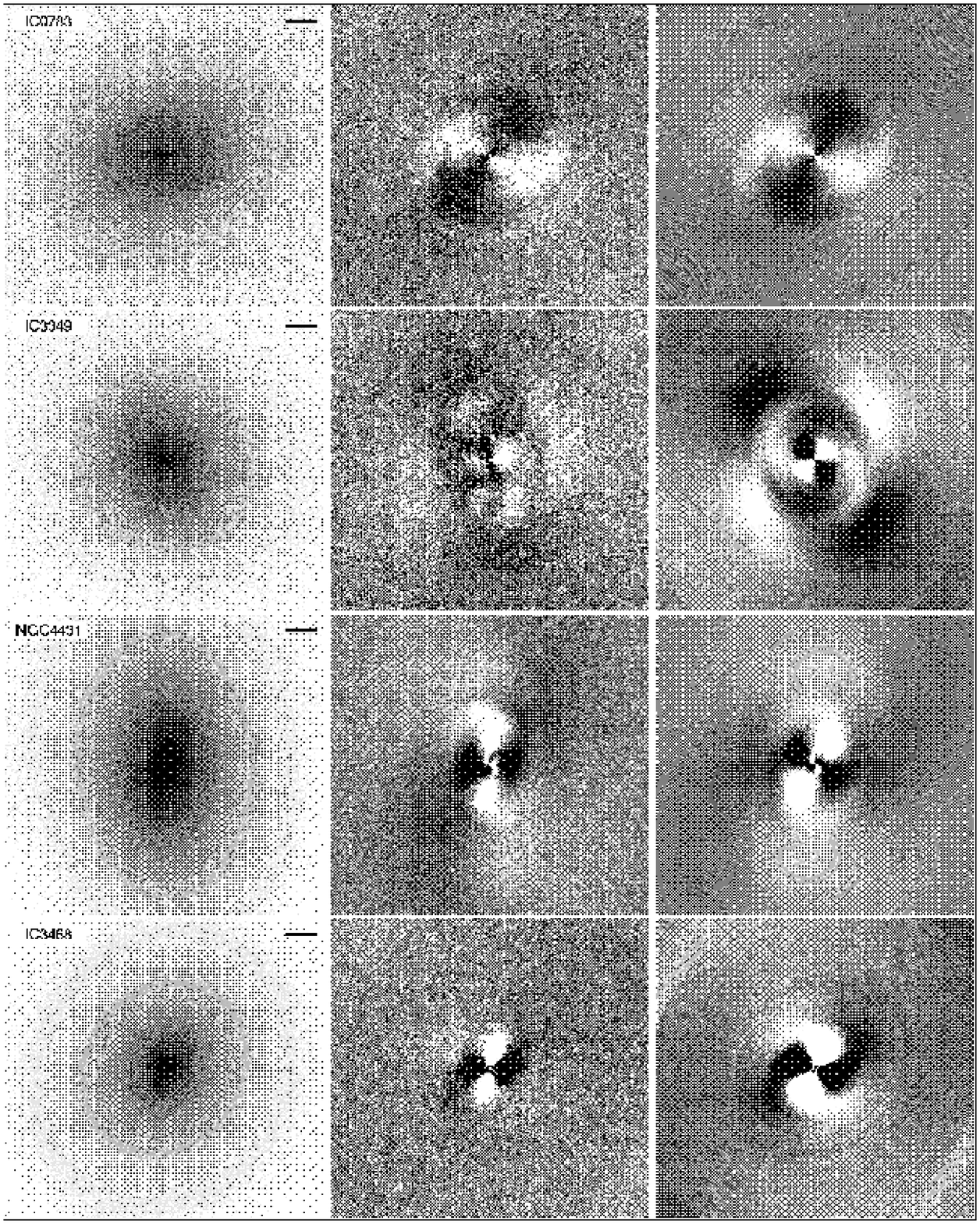,height=192mm,width=144mm}
\end{center}
\caption[]{Left column: $R$-band images of the galaxies. The bar in the
upper right corner corresponds to $10''$ ($\sim 1$ kpc). Middle column:
residuals obtained by subtracting the mean, deprojected surface brightness
profile, corresponding to the first Fourier coefficient $I_0$ from
the $R$-band image. Right column: representation of the third Fourier
coefficient (Fourier map), see text. Image size: $1\farcm7\times 1\farcm7$.
North is up and east to the left.}
\end{figure*}

Direct $R$-band images of the galaxies are shown in the left column of Fig.~1.
The image reduction and the determination of the standard photometric
parameters have been performed with ESO's image processing package
MIDAS. The detailed results of the photometry, like surface brightness profiles,
effective radii, colour gradients etc., as well as best-fitting Sersic
parameters will be presented elsewhere (Barazza
et al., in preparation). Using the MIDAS application FIT/ELL3 we fitted
ellipses to the isophotes of the galaxies and, hence, determined their centers,
ellipticities and position angles.
The ellipticities and position angles of the fitted ellipses versus the
equivalent radius, $r = \sqrt{ab}$, are shown in the top two panels of 
Fig.~2. The variations of these two parameters are indicative of 
the asymmetric features we are interested in.

\section{Residual images and Fourier expansion}
In a first approach we subtracted the deprojected and azimuthally averaged
light distribution of a galaxy from the original image. 
More precisely, we reconstructed
a galaxy using the measured surface brightness profile with {\it fixed}
ellipticity and position angle, as given in cols. 8 and 9 of Table 1, and
subtracted this model from the
original. The results derived in this manner are
shown in the second column of Fig.~1.
Strong residuals are evident in the inner parts of all four galaxies.

\begin{figure*}[t]
\begin{center}
\epsfig{file=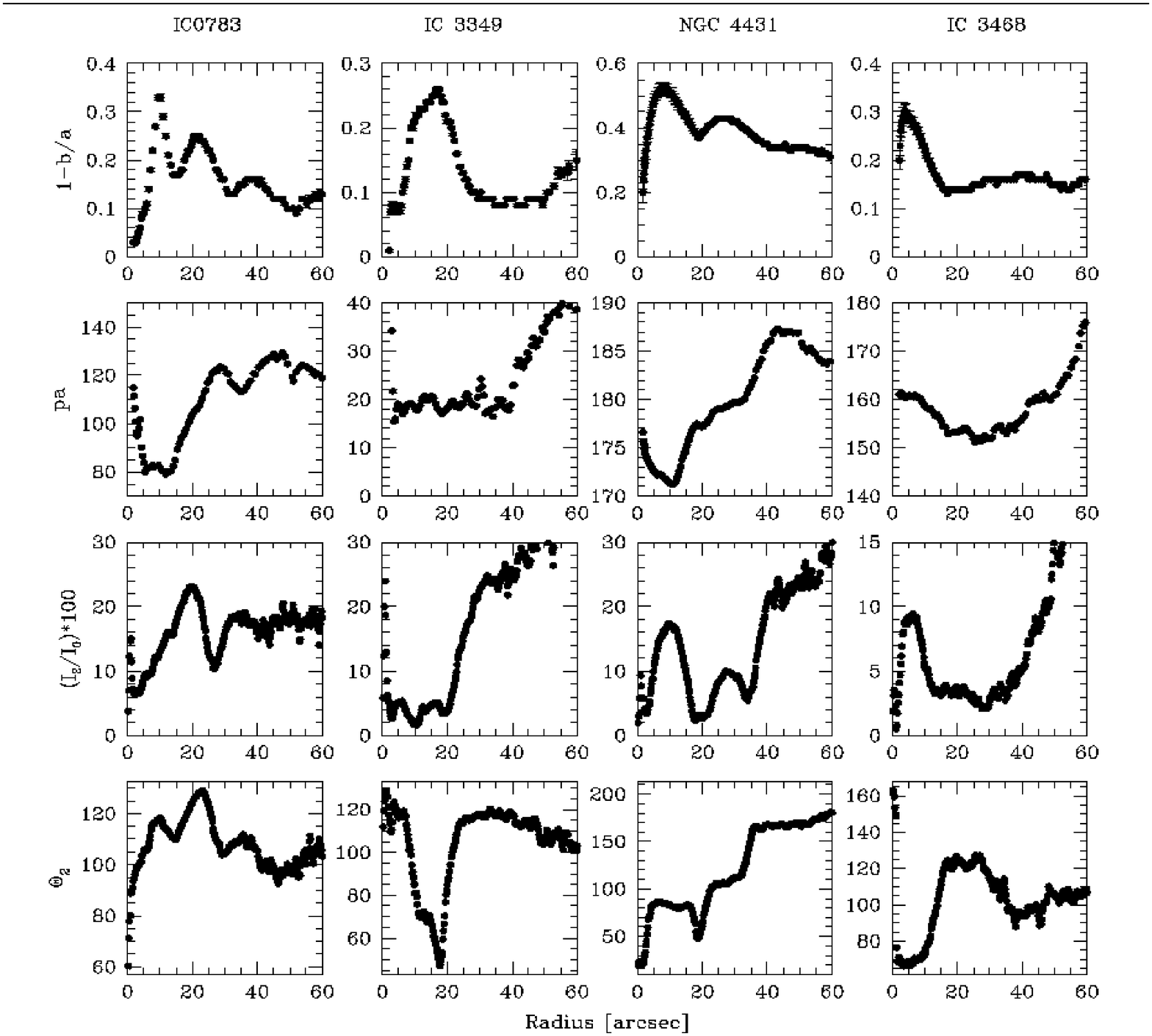,height=150mm,width=150mm}
\end{center}
\caption[]{First row: ellipticity profile; 
second row: position angle profile. Third row: relative strength, in
percent, of the third Fourier coefficient $I_2$ relative to $I_0$. Fourth
row: profile of the phase angle $\Theta_2$ of the third Fourier coefficient.
All parameters are plotted against the equivalent galactocentric radius.}
\end{figure*}

IC0783 and NGC4431 show the most striking features; 
in both cases weak signs of spiral or bar structure can be discerned, or
guessed, already in the direct images (left column).
The spiral arms of IC0783 are 
beautifully traced by the oscillations in the ellipticity and position angle 
profiles plotted in Fig.~2.
The residual of IC3349, which otherwise is a normal, almost round 
dwarf elliptical with a bright nucleus, shows an extended and seemingly lumpy
structure along the major axis. The bright
knot about $10''$ south-east of the center is caused by a luminous
background galaxy. NGC4431 is certainly the most interesting case.
In the northern and southern part of its $R$-band image two dark
arcs are visible (unfortunately barely so in the printed version). 
The residual shows
that these arcs are at the end of a bar, forming the outer edges of a
spiral-like feature, probably the trailing arms of the rotating bar.

Finally, in the residual of IC3468 we note a prominent hourglass-like feature.
This is in fact what is expected to appear 
whenever the ellipticity in the central part
of a galaxy significantly
changes with radius, and it was observed in more than these four galaxies
(cf.~Sect.~1). Indeed, by inspecting Fig.~2 (top row), we see that the 
ellipticity profile shows an inner peak in all four cases.
Moreover, if an ellipticity change is combined with only a 
slight change of position angle (twisting isophotes), 
the residual, as a means of amplification, can additionally 
show (sometimes dramatic) spiral structure. Again we note in Fig.~2 (second row
from top) significant position angle changes in IC0783 and NGC4431, where we
found traces of spiral arms.
As ellipticity changes and twisting isophotes
are quite commonly found in spheroidal galaxies, 
possibly as a result of their intrinsic triaxiality
(e.g.~Kormendy 1982, Binney \& Merrifield 1998), 
it is thus clear that hourglass-like
and spiral-like features found in residual images of the kind shown in Fig.~1
must not, at face value, be taken as indication for the presence of a true bar
or true spiral structure (i.e.~disk structure). However, as mentioned in 
the introduction and as demonstrated in the following section, the residual
features in our four dwarfs clearly persist when scrutinized with an 
unsharp masking method. Their strength makes it difficult to explain these
features as mere artifacts produced by such 
parameter combinations; it is more likely that they indicate the presence of a
real disk (bar and/or spiral) structure that is naturally accompanied 
by a systematic change in ellipticity and position angle when ellipses are 
fitted to the surface brightness distribution.     

To analyse the residual features found in a more quantitative manner, we
then Fourier decomposed the direct images. This technique has frequently 
been used in the
past to quantify non-axisymmetric structures in disk galaxies (Grosbol 1987;
Elmegreen et al. 1989; Rix \& Zaritsky 1995, Vera-Villamizar et al. 1998).
We used the Fourier expansion in the polar form, with azimuthal angle
$\Theta$. To this end we had to determine the centers of the galaxies first.
Since we did not measure a considerable offset for any of the nuclei, we
derived the centers by fitting a Gaussian to them. The pixel
positions could then be expressed in polar coordinates. All isophotes were
subsequently deprojected using the ellipticities and position angles from 
Table 1, and represented by 128 pixels uniformly distributed over each
isophote. Finally, the corresponding intensities, $I(r,\Theta)$, were expanded
in the Fourier series:
\begin{eqnarray}
I(r,\Theta) = I_0(r) + I_1(r)\cos[\Theta-\Theta_1(r)] + \nonumber \\
I_2(r)\cos2[\Theta-\Theta_2(r)] + ....
\end{eqnarray}
The zeroth order Fourier coefficient, $I_0$, only depends on the radius and is
therefore equivalent to the mean surface brightness of the galaxy. Actually,
converting $I_0$ in magnitudes and plotting it versus radius yields the mean
surface brightness profile. The first order Fourier coefficient, $I_1$,
characterises the offset of the isophote with respect to the chosen center.
Hence, this parameter is a measure of the lopsidedness of a galaxy and the
associated phase angle, $\Theta_1$, indicates the direction of the offset.
It could also be used to measure the offset of the nucleus from the
center of the brightness distribution. The next coefficient, $I_2$, is
associated with even-numbered asymmetries, e.g. structures like bars or an even
number of spiral arms. Since the residuals in Fig.~1 show structures of this
kind, we plotted the relative strength of $I_2$, expressed as $(I_2/I_0)*100$,
in the third row of Fig.~2 (intensity profile). The values shown therefore
indicate the strength of the structures in percent of the underlying
luminosity. In the fourth row the corresponding phase angle, $\Theta_2$, is
shown. This angle points to the position of the signal on the
isophote. 
With respect to the radius it remains constant for a bar, indicating
its position angle (e.g.~NGC4431), and rises/decreases linearly for a spiral
pattern (as beautifully seen in IC3328; cf.~Jerjen et al.~2000).

However, the interpretation of this phase angle profile is not always so clear.
We have therefore developed a method to map
the third term of the Fourier expansion, i.e.
$I_2(r)\cos2[\Theta-\Theta_2(r)]$. Since each pixel belongs to an ellipse
around the center, we can expand this specific ellipse in the given Fourier
series and transform the pixel intensity according to
\begin{eqnarray}
I_{x,y}(r,\Theta) = I_2(r)\cos2[\Theta_{x,y}-\Theta_2(r)]
\end{eqnarray}
where $x,y$ are the pixel coordinates. This procedure yields a visual
representation of the information contained in the plots of the third and
fourth row in Fig.~2. In the following, we denote these representations as
``Fourier maps''. The results for the sample galaxies are shown in the third
column of Fig.~1. The Fourier maps are very similar to the residuals (middle
column of Fig. 1), confirming that the structures shown by the residuals are of
the symmetric type, giving a strong signal in $I_2$. 

Let us now go through our sample galaxies, case by case.
Considering the direct image and the ellipticity profile, IC0783 is clearly 
showing
spiral structure. Unfortunately, the residual and the Fourier map, 
as well as the phase
angle profile are not equally revealing. The reason for this might be the
existence of an inner ring: its phase angle would be 
undefined and could cause the observed hump between $10''$ and $15''$.
IC3349 is rather puzzling. The residual image of this galaxy in Fig.~1 
shows an elongated, bar-like structure
which is also evident 
in the ellipticity profile, where it appears as a prominent maximum 
around $20''$, and in the
position angle, which is more or less constant in the same range. 
However, this does not appear in
the Fourier parameter plots. The phase angle is strongly varying
and the intensity
profile shows almost no signal in this region.
In contrast, NGC4431 is a clear case. The phase angle remains 
nearly constant between $4''$ and $17''$,
the $I_2$ profile shows a corresponding luminosity excess of
$\sim 18\%$, and the ellipticity profile is considerably peaked: 
all of this clearly points to the presence of a bar. 
The case of IC3468 is similar to this, but
on a much smaller scale and with lower intensity. A sharp peak in the
ellipticity and intensity profile, along with a constant phase angle are
again pointing to a bar, whose length and intensity reach about half
the values shown by NGC4431.

\begin{figure*}[t]
\begin{center}
\epsfig{file=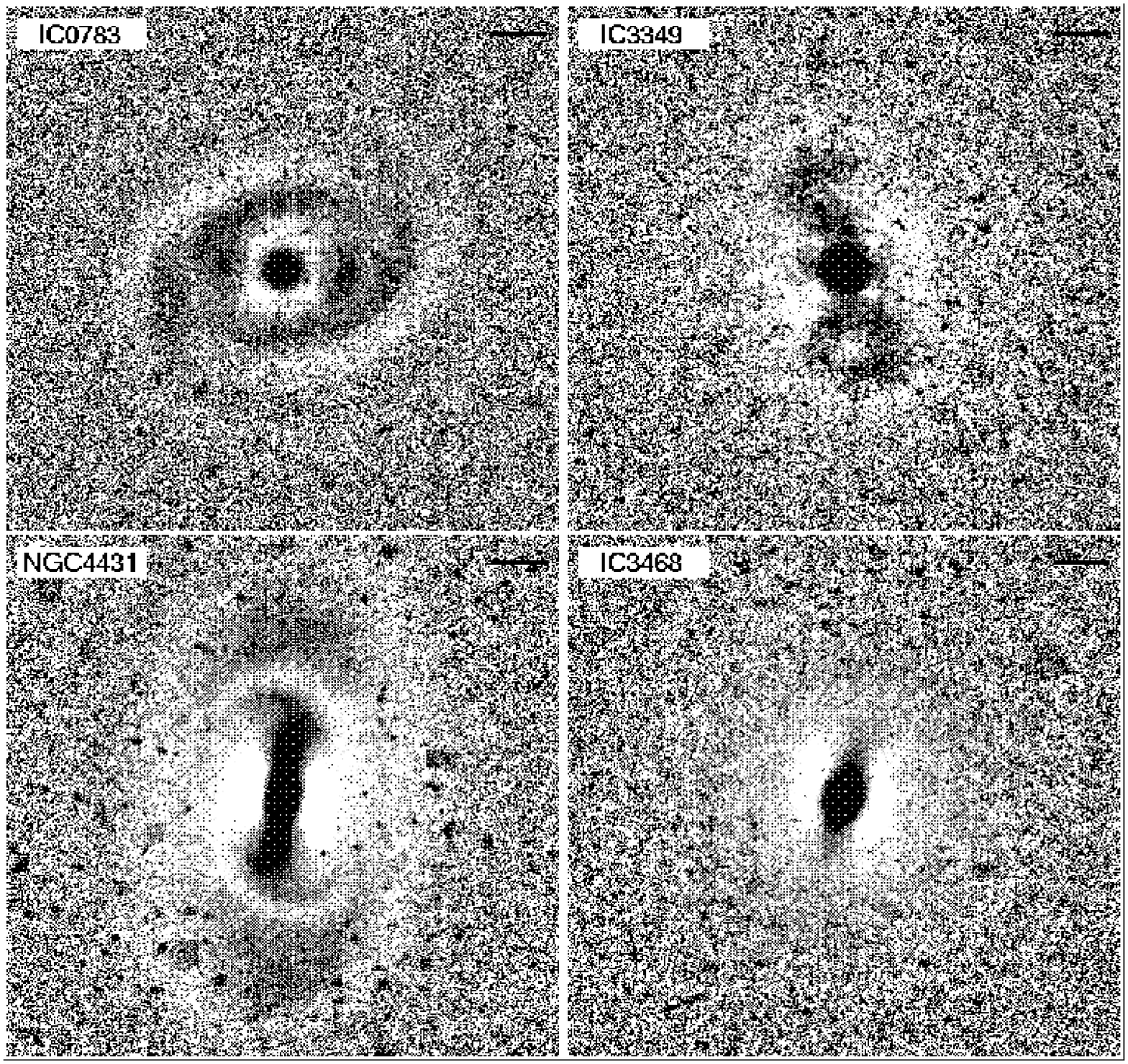,height=150mm,width=150mm}
\end{center}
\caption[]{Result of unsharp masking for the four galaxies. The images are
negatives. Size and orientation as in Fig.~1.}
\end{figure*}

\section{Unsharp Masking}
A well-known method to uncover hidden structures that is independent of
the Fourier analysis, and that also has the advantage that no assumptions are
required, is unsharp masking. This method, which is
also applicable to photographic
plates (Malin \& Zealey 1979), was widely used in the search for hidden inner
structures in different objects. It was successful in investigating the fine
structures in elliptical galaxies (Schweizer \& Ford 1985), dust features in
globular clusters (Mendez et al. 1989), and the true nature of bars in disk
galaxies (Buta \& Crocker 1993). Recently, Colbert et al. (2001) investigated
the morphological differences between elliptical galaxies in different
environments using, among other methods, unsharp masking.

The method essentially consists of deconvolving the optical image with an
appropriate function. Often a point spread function, like a Gaussian, is used.
We chose an even simpler approach: we produced a smoothed image by 
replacing each pixel intensity with the average intensity in a certain 
area around that pixel. The optimal 
size of this area obviously depends on the extension, or scale of the 
structure to be uncovered. We fixed this size simply by trying different
smoothing lengths until the result was satisfactory. Applying the MIDAS task
FILTER/SMOOTH, we obtained the best results with a smoothing radius of $30$
pixels, corresponding to $6''$. Hence, each pixel intensity 
was set to the average
intensity in a square of $31 \times 31$ pixels. The original
image was then divided by this smoothed version. The result of this procedure
for our four galaxies is shown in Fig.~3. 

All four objects show very strong residual features -- we will discuss them
in detail in the following section. What was guessed before
on the basis of ``global'' residual images and a Fourier analysis now
appears with utmost clarity. It
is important to note again that {\it no assumptions}\/ are required here. We
do not have to assume a global ellipticity and position angle in advance, 
to which the residuals refer -- with unsharp masking we determine the residual
light with respect to the {\em local}\/ environment based on the observed 
brightness distribution alone. 

Whenever residual features are uncovered in otherwise perfectly
spheroidal galaxies, the presence of dust has to be considered as a possible
cause (dust lanes in the cores of bright ellipticals are quite common; 
cf.,e.g., Binney \& Merrifield 1998). Dust can be detected by its 
reddening of the colour. However, the $B-R$ colour maps of our galaxies do not
show any conspicuous features other than a slight symmetric colour gradient
(to be reported elsewhere). The only possible exception is NGC4431 whose
colour map reveals that the bar may be slightly redder than the
rest of the galaxy. In particular, IC3349, where dust could be
suspected because
the bar-like feature appears somewhat lumpy, is perfectly 
smooth in colour as well. This means that the residual brightness
features reflect density enhancements in the distribution of stars -- which 
of course is the case with true bars and spirals.

\section{Discussion}
The disk nature of {\bf IC0783} has been conjectured before and this object has
therefore been classified as dS0 (Sandage \& Binggeli 1984; Binggeli \&
Cameron 1991). The possible presence of two spiral arms has already been
reported (Barazza et al. 2001; Jerjen et al. 2001). This
finding is clearly confirmed by our results. To what kind of central structure
these arms are connected is not so evident, however. Since the phase angle
is not constant in the central part, the presence of a bar is not very likely.
We rather suggest the existence of an inner ring, as mentioned in Sect.~3. 
The strongest luminosity excess is observed at the
locations where the spiral arms start, i.e.~the connection points between the
ring and the arms. This can be seen in the Fourier maps and also in the
intensity profile, which reaches a relative excess of $\sim 24\%$. 

During the refereeing stage of this paper, Simien \& Prugniel (2002) have
published a rotation curve of IC0783 (the authors classify this
galaxy as SAB(rs)0/a in the de\,Vaucouleurs system, which is in agreement with
our own new classification). Not surprisingly, as judged from the spiral
morphology,
the galaxy does show rotation, though only with a observed maximal
radial velocity of $v_{max}$ = 22 km\,s$^{-1}$, or $v_{rot}$ =
33 km\,s$^{-1}$ in the plane
of the disk, 
if we assume that the galaxy's apparent ellipticity of 0.25 (Table 1) 
reflects the disk's inclination only. The observed central velocity
dispersion is even a bit larger with $\sigma$ = 36 km\,s$^{-1}$ (however, the 
uncertainties, also for the inclination, are considerable).    

IC0783 is
situated in a region of rather low density within the Virgo cluster. The
distance to the cluster center, i.e.~M87, is quite large. In fact, this galaxy
is probably a member of a group around the giant Sbc galaxy M100. The projected
distance between M100 and IC0783 is $\sim 100$ kpc and the radial heliocentric
velocities are $v_{\odot} = 1571$ km s$^{-1}$ and $v_{\odot} = 1293$ km
s$^{-1}$ for M100 and IC0783, respectively (all velocities from the NED).
IC0783 could therefore be bound to
M100. The whole group is believed to have fallen into the cluster recently
(Binggeli et al. 1987).
Even if the group were falling in for the first time, IC0783 could
already have been stripped of its gas and stopped star formation, in accord
with its rather red colour of $B-R = 1.34$ and the non-detection of the
spiral structure in the colour map.

{\bf IC3349} has been classified as dE1,N; it is very round and has a bright
nucleus. The image from unsharp masking reveals an elongated structure almost
along the major axis of the galaxy. As mentioned in Sect.~3, the knot lying
south-east of the nucleus is caused by a background galaxy. Therefore, the
feature is quite symmetric. The
best photometric evidence for the structure is provided by the
ellipticity profile, which shows a pronounced maximum at $\sim 20''$.
Despite its weakness (the luminosity excess amounts only to $\sim 5\%$),
the structure clearly resembles a bar. Its thick, dumb-bell-like ends
are in fact reminiscent of the classical SB0$_3$ Hubble type. There is even
a slight hint for rudimentary spiral/trailing arms as in NGC4431
(though much less clearly). In any case,
the presence of a bar suggests that IC3349 is a rotating disk galaxy. There
is no kinematic data available yet to confirm this. IC3349 is
located in the Virgo M87 subcluster as well, i.e. in the densest region of the
cluster. It is more or less between M87 and M86, but closer to the latter,
and does not seem to be associated with a giant galaxy.

The most impressive discovery is the beautiful bar in {\bf NGC4431}. 
This galaxy has been classified as dS0 before,
though without acknowledging the presence of a bar (Sandage \& Binggeli 1984);
we now would call it a dSB0/a. Its twisting
isophotes have been pointed out by Ryden et al.~(1999), where a contour
plot of a $V$-band image is shown. Also Gavazzi et al.~(2001) observed
this galaxy in the $B$- and $V$-band and performed a profile decomposition.

In the unsharp masked image in Fig.~3 the bar and the trailing
arms of NGC4431 are very striking. The whole morphology is reminiscent
of the results of a bar simulation
presented by Patsis \& Athanassoula (2000). The gas density distribution 
plotted in Fig.~5 of that paper shows exactly the same features. 
In particular, the authors draw 
attention to the typical ``T-structure'', consisting of the bar, the 
trailing arms plus
short density enhancements on the leading side of the bar. 
Even the gaps beyond the outer edges of the trailing arms, followed by
another pair of denser filaments, are much the same in NGC4431 and 
the simulation. Given this similarity, we can assume that the bar
in NGC4431 is rotating clockwise. However, we should also note a 
difference in size: the bar in the simulation has a length of $\sim 10$ kpc,
whereas the observed one measures only $\sim 3$ kpc. But just this
smallness may provide another interesting aspect: As dwarf galaxies are
believed to contain large amounts of dark matter, and as the pattern speed
of a bar is governed by the mass distribution via dynamical friction
(Debattista \& Sellwood 1998, 2000), objects like NGC4431 might offer
new possibilities to put constraints on the dark matter content of dwarf
galaxies.

The pattern speed of the bar in NGC4431 is still unknown, but its
solid body rotation has been measured: the rotation curve determined by Simien
\& Prugniel (2002) nicely shows the expected linearity over the scale of the
bar. The maximum rotational velocity for NGC4431 is listed as $v_{max}$ = 60
km s$^{-1}$. As the galaxy is apparently strongly flattened ($\epsilon$ =
0.38, Table 1), meaning its disk
is strongly inclined, this transforms into a mere $v_{rot}$ = 76 km s$^{-1}$
in the plane of the disk -- barely larger than the measured central velocity
dispersion of $\sigma$ = 55 km s$^{-1}$. 

NGC4431 is close to the center of the Virgo cluster; its projected distance
to M87 is $\sim 300$ kpc. It is therefore located in a rather
dense region and belongs to subcluster A (Binggeli et al. 1987). In fact
the dwarf-like SBa galaxy NGC4440 and another dS0 type, NGC4436, 
lie within a projected radius of
$\sim 40$ kpc (see image in Sandage \& Binggeli 1984). However, the radial
velocities of these galaxies
are quite different. NGC4431 itself has $v_{\odot}
= 913$ km s$^{-1}$, the closest neighbors (in projection), NGC4436 and NGC4440,
have $v_{\odot} = 1163$ km s$^{-1}$ and $v_{\odot} = 724$ km s$^{-1}$,
respectively. A physical grouping of these dwarfs is therefore unlikely

Finally, a small, bar-like feature is evident in the unsharp masked image 
of {\bf IC3468}. The peak in the ellipticity profile is very narrow and the 
relative
intensity reaches almost $\sim 10\%$. Note that 
the galaxy has been classified E1,N instead of dE1,N -- precisely on the basis
of the high-surface brightness central core that was found to be indicative
of a classical (non-dwarf) elliptical. 
It is not clear whether we see a small
disk nearly edge-on embedded in an almost round (dwarf) elliptical, 
or a very short bar embedded in a disk galaxy almost face-on. 
Judging from the 
pointed (lemon-shaped) isophotes of the residual feature (as 
confirmed by a ``disky'' value of $a_4/a \sim$ 0.01 in the inner 15''; 
see below), a disk component within a spheroid would seem more likely. 
Surprisingly, however, Simien \& Prugniel (2002) were not able to measure
any appreciable rotation in IC3468, listing $v_{max} < 9$ km s$^{-1}$ for
this galaxy (along the correct position angle of the elongated inner feature).
An embedded edge-on disk can therefore be excluded. But a bar
in an almost face-on disk is a difficult interpretation as well, because 
with $\epsilon$ = 0.17 (Table 1) the deprojected velocity 
of $v_{rot} < 16$ km s$^{-1}$ would still be 
much smaller than the central velocity dispersion $\sigma$ = 36 km s$^{-1}$.
With $v_{max}/\sigma$ = 0.25, IC3468 appears to be a typical anisotropic
(non-rotation-supported) dwarf elliptical. 
One might argue, though, that the
maximum of the rotation curve was not reached by Simien \& Prugniel 
(2002); their measurements indeed cover only the innermost (barred) part of
the galaxy. We think the case -- for or against the presence of disk structure
in this galaxy -- is left open.   
As to its position in the cluster, IC3468
is situated between the A subcluster, dominated by M87, and the M49 subcluster
in the south of Virgo and is therefore in a region of rather low
density.

We have also done an isophotal analysis for all our Virgo dwarfs observed
with the VLT, the results of which will be reported elsewhere (Barazza et al.,
in preparation). In such an analysis one is looking for systematic deviations
of the isophotes 
from pure ellipses, again by means of a Fourier method.
The fourth Fourier coefficient, $a_4$, determines whether
the isophotes are ``disky'' ($a_4>0$) or ``boxy'' ($a_4<0$). 
This parameter is believed
to be sensitive to weak disk or bar components in elliptical galaxies,
depending on the
inclination angle (Carter 1987; Bender \& M\"ollenhoff 1987;
Rix \& White 1990; Ryden et al. 1999).
Our disk/bar dwarf candidates do show some correlation with the $a_4$
parameter. We have already mentioned the inner diskyness of IC3468.
While IC0783 does not show a coherent $a_4$ signal, NGC4431 is slightly boxy,
as is IC3349 slightly disky (surprisingly), in the central residual
region.
However, in general we think the meaning of $a_4$ is difficult to interpret;
different combinations of components can produce the same $a_4$ profile.
In particular, disky isophotes need not necessarily mean that there is a 
true disk component embedded in the galaxy.
We are currently exploring the principal possibilities to model  
$a_4$ profiles.


The best evidence for the presence of disk components in dwarf ellipticals or
dS0s can be expected to come from kinematic studies. These difficult 
observations are now
underway for a large number of early-type 
dwarfs in the Fornax and Virgo clusters 
(de Rijcke et al. 2001; Geha et al. 2001). So far, the 
majority of dwarfs have turned out
to be non-rotating spheroids. However, a few are rotation-supported and may
plainly be disk galaxies (Simien \& Prugniel 1998). Interestingly, this holds
only for Virgo cluster dwarfs; no rotators were found so far among Fornax
dEs, nor did any of them exhibit the kind of spiral or bar features reported
in the present paper (de Rijcke, private communication). 

Our findings support the conjecture that a rather large number of bright,
early-type dwarf galaxies, at least in the Virgo cluster, 
are disks or at least possess a disk component. These
systems might formerly have been late-type disk galaxies which, by
interactions with the cluster, were transformed to the systems we observe
today. Such processes are described for example in the harassment scenario
developed by Moore et al.~(1998). In this picture, galaxies falling into the
cluster lose all of their gas and most probably 
transform to a spheroid by gravitational interactions with the other cluster
members. However, massive late-type galaxies might preserve their disk nature,
at least for some time. The galaxies discussed could therefore be 
transient remnants of the harassment process. Also with respect to this
scenario, the Fornax cluster may be a different galaxy environment because it
is more isolated in space. The Virgo cluster is surrounded by clouds of
late-type galaxies that seem to be fed into the cluster at a constant rate
(see, e.g., Binggeli et al.~1987). Perhaps the high frequency of hidden disks
in dEs is unique for the Virgo cluster environment. 

\section{Summary and Conclusions}
Following the discovery of spiral structure in a early-type dwarf galaxy
classified as dE,N by Jerjen et al.~(2000), we searched our whole VLT-sample of
bright Virgo dEs and dS0s for further photometric disk signatures. 
The principal search tools applied were (1) looking for inner residual 
features by subtracting a model image of the galaxy
based on its mean surface
brightness profile, but with fixed ellipticity and position angle,
from the observed image; (2) a Fourier expansion of the galaxy image
in polar form, where the lowest order of non-axisymmetry is indicating
the presence of a bar or two-armed spiral structure; and (3) unsharp masking,
where the original image is divided by an appropriately smoothed one
to enhance any local ``irregularities''.
Unsharp masking turned out to be the most reliable method
to uncover hidden structures. The first method can give misleading results,
if the inherent assumptions on global symmetry, as well as the high
sensitivity of the outcome to slightly varying position angles
are not taken into account. But once the clear-cut choice by unsharp masking
is made, the first two methods are useful to visualize and quantify the   
symmetric residual structures found.

In addition to IC3328 (Jerjen et al.~2000), we found photometric traces
of a possible disk component in four more early-type dwarfs out of a
sample of 23:

{\bf IC0783}: The two spiral arms of this ``dS0'' galaxy are
already evident in the direct optical image. Obviously this is a disk galaxy,
which now is also confirmed by the measurement of its rotation
(Simien \& Prugniel 2002). The central structure of this galaxy remains
unresolved; we think there could be an inner ring. 

{\bf IC3349}: Fourier analysis and unsharp masking
reveal a long and elongated (if only weak) structure in the central
part, which we interpret as a bar in a 
nearly face-on disk.
A revised type for this ``dE'' galaxy would be dSB0.

{\bf NGC4431}: The quite strong bar present in this galaxy is the most
striking discovery and clearly reveals the disk nature of this dwarf --
again nicely confirmed by its measured rotation (Simien \& Prugniel 2002). 
Besides
the bar we clearly note trailing arms and two dense regions on the leading
side of the bar. This so-called T-structure is very similar to the results 
of a simulation presented by Patsis \& Athanassoula (2000). 
A more fitting type for this ``dS0'' galaxy would be dSB0/a.

{\bf IC3468}: In the very center of this dwarf elliptical we either observe a
rather short bar in a nearly face-on disk, or a small disk seen edge-on in a
spheroid. Surprisingly, Simien \& Prugniel (2002) found essentially
zero rotation along the position angle of this structure,
which renders a clear interpretation of what we see impossible at present. 

We emphasize that none of these objects is
comparable to IC3328. The weak and uniformly wound spiral structure in this
galaxy seems to be truly unique, constituting a particular 
class of dwarf galaxies; at least we did not find an additional example.

Our findings confirm previous suggestions that a sizeable fraction of 
all bright early-type dwarfs in
the Virgo cluster are disk galaxies. In a possible
scenario for their evolution they are former late-type disk galaxies which
have been transformed to the systems we observe today during their infall to
the cluster. The discovery of more objects of this kind in the Virgo cluster,
but also in other clusters, could therefore further constrain possible models
for the formation and evolution of early-type galaxies in general.

\begin{acknowledgements}
The authors thank Victor Debattista for inspiring discussions. F.D.B and B.B.
are grateful to the Swiss National Science Foundation for financial support.
This research has made use of the NASA/IPAC Extragalactic Database (NED) which
is operated by the Jet Propulsion Laboratory, California Institute of
Technology, under contract with the National Aeronautics and Space
Administration, as well as NASA's Astrophysical Data System Abstract Service.
\end{acknowledgements}


\begin{thebibliography}{}

\bibitem{}
Abadi, M.G., Moore, B., Bower, R.G. 1999, MNRAS, 308, 947

\bibitem{}
Barazza, F.D., Jerjen, H., Binggeli, B. in preparation

\bibitem{}
Barazza, F.D., Binggeli, B., Jerjen, H. 2001, in ``Dwarf Galaxies and
their Environment'', eds. K.S. de Boer, R.-J. Dettmar, U. Klein, Shaker
Verlag, p. 243

\bibitem{}
Bender, R., M\"ollenhoff, C. 1987, A\&A, 177, 71

\bibitem{}
Binggeli, B., Sandage, A., Tammann, G.A. 1985, AJ, 90, 1681

\bibitem{}
Binggeli, B., Tammann, G.A., Sandage, A. 1987, AJ, 94, 251

\bibitem{}
Binggeli, B., Cameron, L.M. 1991, A\&A, 252, 27

\bibitem{}
Binggeli, B., Popescu, C.C. 1995, A\&A, 298, 63

\bibitem{}
Binney, J., Merrifield, M. 1998, ``Galactic Astronomy'', Princeton University
Press

\bibitem{}
Buta, R., Crocker, D.A. 1993, AJ, 106, 939

\bibitem{}
Carter, D. 1987, ApJ, 312, 514

\bibitem{}
Colbert, J.W., Mulchaey, J.S., Zabludoff, A.I. 2001, AJ, 121, 808

\bibitem{}
Debattista, V.P., Sellwood, J.A. 1998, ApJ, 493, L5

\bibitem{}
Debattista, V.P., Sellwood, J.A. 2000, ApJ, 543, 704

\bibitem{}
De Rijcke, S., Dejonghe H., Zeilinger, W.W., Hau, G.K.T. 2001, ApJ, 559, 21

\bibitem{}
Elmegreen, B.G., Elmegreen, D.M., Seiden, P. 1989, ApJ, 343, 602

\bibitem{}
Ferguson, H.C., Binggeli, B. 1994, A\&AR, 6, 67

\bibitem{}
Gavazzi, G., Zibetti, S., Boselli, A., Franzetti, P., Scodeggio, M., Martocchi,
S. 2001, A\&A, 372, 29

\bibitem{}
Geha, M., Guhathakurta, P., van der Marel, R. 2001,
in ``The Shapes of Galaxies and their Halos'', Yale Cosmology Workshop,
ed. P. Natarjan

\bibitem{}
Grosbol, P. 1987, in ``Selected Topics on Data Analysis in Astronomy'', ed. L.
Scarsi, V. DiGesu \& P. Crane, World Scientific, p. 57


\bibitem{}
Jerjen, H., Binggeli, B., Kalnajs, A. 2000, A\&A, 358, 845

\bibitem{}
Jerjen, H., Kalnajs, A., Binggeli, B. 2001, in ``Galaxy Disks and Disk
Galaxies'', eds. Jos\'e G. Funes, S. J. and Enrico Maria Corsini, ASP
Conference Series, Vol. 230, p. 239

\bibitem{}
Jerjen, H., Binggeli, B., Barazza, F.D. in preparation

\bibitem{}
Kormendy, J. 1982, in ``Morphology and Dynamics of Galaxies'', 12th Advanced
Saas-Fee course, eds. J. Binney, J. Kormendy, S.D.M. White, Swiss Society
for Astrophysics and Astronomy, p. 149

\bibitem{}
Malin, D.F., Zealey, W.J. 1979, S\&T, 57, 354

\bibitem{}
Mendez, M., Orsatti, A.M., Forte, J.C. 1989, ApJ, 338, 136

\bibitem{}
Moore, B., Lake, G., Katz, N. 1998, ApJ, 495, 139

\bibitem{}
Nieto, J.-L., Bender, R., Poulain, P., Surma, P. 1992, A\&A, 257, 97

\bibitem{}
Patsis, P.A., Athanassoula, E. 2000, A\&A, 358, 45


\bibitem{}
Rix, H.-W., White, S.D.M. 1990, ApJ, 362, 52

\bibitem{}
Rix, H.-W., Zaritsky, D. 1995, ApJ, 447, 82

\bibitem{}
Ryden, B.S., Terndrup, D.M. 1994, ApJ, 425, 43

\bibitem{}
Ryden, B.S., Terndrup, D.M., Pogge, R.W., Lauer, T.R. 1999, ApJ, 517, 650


\bibitem{}
Sandage, A., Binggeli, B. 1984, AJ, 89, 919

\bibitem{}
Schweizer, F., Ford, W.K. Jr. 1985, in ``New Aspects of Galaxy
Photometry'', ed. J.L. Nieto, Springer Verlag, p. 145

\bibitem{}
Simien, F., Prugniel, Ph. 2002, A\&A, 384, 371



\bibitem{}
Vera-Villamizar, N., Dottori, H., de Carvalho, R., Puerari, I. 1998, in
``Focal Points in Latin American Astronomy'', eds. Aguilar, A., Carraminana,
A., Revista Mexicana de Astronomia y Astrofisica Serie de Conferencias.

\end{thebibliography}
\end{document}